\documentclass[traditabstract]{aa} % for a referee version
\usepackage{graphicx}
\usepackage{natbib}
\usepackage{amsmath}
\usepackage{mwe,tikz}
\usepackage[percent]{overpic}

\usepackage[para,online,flushleft]{threeparttable}
\usepackage{txfonts}
\def \hcm {\hbox {\ifmmode $ atom cm$^{-2}\else atom cm$^{-2}$\fi}}

\def \apjl {ApJL}
\def \apjs {ApJS}\def \aap {A\&A}

\begin{document}
   \title{Clumpy wind accretion in supergiant neutron star high mass X-ray binaries}

   \author{E. Bozzo
    \inst{1}
    \and L. Oskinova
    \inst{2}
   \and A. Feldmeier
     \inst{2} 
   \and M. Falanga
     \inst{3,4}            
    }

   \institute{ISDC Data Centre for Astrophysics, Chemin d’Ecogia 16,
    CH-1290 Versoix, Switzerland; \email{enrico.bozzo@unige.ch}
    \and   
    Institut f\"ur Physik und Astronomie, Universit\"at Potsdam,
   Karl-Liebknecht-Strasse 24/25, 14476 Potsdam, Germany 
     \and   
	 International Space Science Institute (ISSI), Hallerstrasse 6, CH-3012 Bern, Switzerland
     \and   
	 International Space Science Institute in Beijing, No. 1 Nan Er Tiao, Zhong Guan Cun, Beijing 100190, China
     }

   \date{}

  \abstract{The accretion of the stellar wind material by a compact object represents the main mechanism 
  powering the X-ray emission in classical supergiant high mass X-ray binaries and supergiant fast X-ray transients. 
  In this work we present the first attempt to simulate the accretion process of a fast and dense massive 
  star wind onto a neutron star, taking into account the effects of the 
  centrifugal and magnetic inhibition of accretion (``gating'') due to the spin  
  and magnetic field of the compact object. We made use of a radiative hydrodynamical code to 
  model the non-stationary radiatively driven wind of an O-B supergiant star and then place a neutron star characterized 
  by a fixed magnetic field and spin period at a certain distance from the massive companion. 
  Our calculations follow, as a function of time (on a total time scale of several hours), the transition of the system through 
  all different accretion regimes that are triggered by the intrinsic variations in the density and velocity of the non-stationary wind. 
  The X-ray luminosity released by the system is computed at each 
  time step by taking into account the relevant physical processes occurring in the different accretion regimes. Synthetic lightcurves are 
  derived and qualitatively compared with those observed from classical supergiant high mass X-ray binaries and supergiant fast X-ray transients. 
  Although a number of simplifications are assumed in these calculations, we show that taking into account the effects of the centrifugal 
  and magnetic inhibition of accretion significantly reduces the average X-ray luminosity expected for any neutron star wind-fed binary. 
  The present model calculations suggest that long spin periods and stronger magnetic fields are favoured in order to reproduce the 
  peculiar behavior of supergiant fast X-ray transients in the X-ray domain.}   
  

   \maketitle

\section{Introduction}
\label{sec:intro}

Most of the supergiant high mass X-ray binaries (SGXBs), including the Supergiant Fast X-ray Transients (SFXTs),  
host a neutron star (NS) accreting from the fast and dense wind of a massive O-B supergiant companion. 
The accretion of the stellar wind material onto the compact object power a conspicuous X-ray 
emission \citep[see][for a recent review]{walter2015}. 

The so called ``classical'' SGXBs are typically characterized 
by a relatively constant average X-ray luminosity. Depending mostly on the 
orbital separation between the NS and the companion, this can range from 10$^{34}$~erg~s$^{-1}$ to 10$^{37}$~erg~s$^{-1}$ and 
display a short term variability achieving a dynamic range in the X-ray luminosity of $10-100$ over timescales of 
hundred to thousand seconds. 
The SFXTs are known, instead, to display a much more extreme behavior in the X-ray domain.  
These sources remain in quiescence 
($L_{\rm X}\sim10^{32}-10^{33}$~erg~s$^{-1}$) for most of the time and only sporadically undergo short and bright outbursts, 
reaching a luminosity comparable to that of the classical systems ($L_{\rm X}\sim10^{36}-10^{37}$~erg~s$^{-1}$). 
The dynamical range of the SFXT X-ray luminosity can thus achieve values as large as $\Delta L_{\rm X}$$\sim$10$^5$-10$^6$.  
The outbursts of these sources last for a few hours at the most, and thus their activity 
duty cycle is estimated to be typically of a few \% \citep{lutovinov13,romano11,romano14,paizis14,bozzo15}. 
In virtually all SFXTs it was also observed that X-ray flares reaching a luminosity of 1-10\% of that of the brightest 
outbursts can occur at anytime, displaying timing and spectral properties similar to those of the more luminous events  
\citep[see, e.g.,][]{sidoli08,bozzo10,bodaghee10,romano13,romano14b}.  

The spectroscopic properties of the SFXT X-ray emission is typical of wind accreting NS systems, and can be usually described by using 
a power-law model with a cut-off around 10-30~keV \citep[see, e.g.,][]{romano08,sidoli09,romano15}. Soft thermal components have been detected 
in the X-ray spectra of the SFXTs both during outbursts and quiescence. In the first case, these components 
are mostly ascribed to hot spots on the NS surface, while in quiescence the thermal emission could be more easily explained 
as being produced within the wind of the supergiant companion \citep{bozzo10,sidoli10b}. 

The winds of OB supergiants are indeed well known to produce a relatively bright X-ray emission that could reach 
values of 10$^{34}$~erg~s$^{-1}$ in the most extreme cases. The origin of these X-rays is still highly debated, but one of the 
most credited hypothesis is that they are produced due to the collision of ``clumps'' in the stellar wind, i.e. structures 
characterized by a higher density and a different velocity compared to the surrounding intra-clump medium 
\citep[see, e.g.,][and references therein]{feldmeier1995,Oskinova2006}. 
One of the stellar wind models developed to study the formation of clumps and the release of X-ray radiation from their collisions 
was presented by \citet{feldmeier97}. This radiatively non-stationary wind model allows us to carry out an hydrodynamic 
investigation of the wind formation and evolution from the first principles, together with the computation of the thermal structure 
of the wind and the energy distribution of the emitted radiation. The model uses a 1D approach and thus the clumps can grow significantly 
in size and density due to the lack of the other two spatial dimensions that would allow us to observe the lateral  
break-up of these structures as a consequence of the Rayleigh-Taylor or thin-shell instability  
\citep{dessart03,dessart05}. Multi-dimensional simulations (mostly 2D and pseudo 3D), however, were so far unable to account for the thermal 
structure of the wind and thus they could not make any clear prediction on the intensity of the produced X-ray radiation to be compared with 
observations of isolated OB supergiants and massive stars in SGXBs \citep[see, e.g.,][]{puls08}.  
More recent findings seem to suggest that clumps might likely be limited to a density ratio of $\sim$10 compared to the 
surrounding medium and a lateral extent of the order of 10\% of the supergiant star radius 
\citep[see, e.g.,][and references therein]{surlan13}.  

The possibility of having massive and dense structures in the winds of supergiant stars, as predicted by the 1D approach of stellar wind models  
mentioned above, stimulated the idea that the SFXTs might have been a sub-class of SGXBs hosting massive stars with extreme  
clumpy winds \citep{zand05,negueruela2006,walter07,negueruela2007,bozzo11}. The accretion 
of a massive clump onto the NS would produce in this scenario bright and short X-ray outbursts, as the duration of the event is 
related to the lateral size of the clump while the total luminosity is directly proportional to its mass. 
Less luminous flares can be explained by invoking a reasonable distribution in mass and size of the clumps \citep{zand05,walter07}. 
The quiescent emission of the SFXT can be interpreted in this scenario by assuming that in this case the  
NS is accreting through the much more rarefied intra-clump medium, which particularly low density can only provide  
a feeble mass accretion rate onto the 
compact object. 

The first attempt to combine non-stationary stellar wind codes with calculations 
on the expected accretion luminosity produced by a NS located inside the wind was presented by \citet{oskinova2012}. 
These authors assumed the simplest accreting scenario in which all the stellar wind material gravitationally captured by the NS is accreted 
onto the compact object. Their calculations showed that the extremely clumpy wind produced by the 1D code gives rise to 
a large X-ray variability reaching a dynamic range of 10$^6$-10$^7$. Even though this is compatible with the dynamic range displayed 
by the SFXTs, the overall behaviour observed in the simulations of \citet{oskinova2012} did not resemble that typically observed from the 
sources in this class. In particular, the model failed to reproduce the SFXT low duty cycle, as bright outbursts were predicted  
to repeat irregularly at any time without the presence of extended periods of quiescence. 

As discussed by \citet{bozzo15}, it is likely that extremely clumpy winds are not enough to explain the complex phenomenology 
of the SFXTs: some additional mechanism is needed to halt the accretion onto the NS for a large fraction of time and 
reproduce the low duty cycles of these systems \citep[see also][]{romano14,sidoli16}.  
\citet[][hereafter B08]{bozzo08} proposed that the inhibition of accretion could 
occur in the SFXTs due to a peculiarly intense magnetic field and slow rotation of the compact 
object\footnote{Note that also other mechanisms have been proposed to inhibite the accretion in the SFXTs 
and produce sporadic bright outbursts \citep[see][]{shakura14}.}. 
The NS rotation and magnetic field were indeed shown to give rise to a centrifugal and magnetic gate that can inhibit 
accretion through either the on-set of a propeller effect \citep{illarionov75aa,grebenev07} or preventing the gravitational 
focusing of the wind material toward the NS. 

In this paper, we made a step forward compared to the simulations presented by \citet{oskinova2012} and show how 
the magnetic and centrifugal gating mechanisms proposed by B08 would work in presence of a non-stationary clumpy stellar wind. 
Our aim is to provide a refined calculation of the accretion luminosity that arise in wind accreting systems and investigate 
on time scales spanning several hours if the centrifugal and magnetic gating mechanisms can help reproducing the characteristic 
behaviour of the SFXTs in the X-ray domain. In Sect.~\ref{sec:clumpywind} we summarize the main properties of 
the 1D clumpy wind model already used by \citet{oskinova2012}, and provide a brief description of the gating accretion models of 
B08 in Sect.~\ref{sec:gating}. The results obtained by taking into account the gating mechanisms during the accretion of the 
clumpy stellar wind material onto a NS are described in Sect.~\ref{sec:results}. 
We provide our discussion and conclusions in Sect.~\ref{sec:conclusions}.   
\begin{figure*}
\centering
\includegraphics[width=15cm]{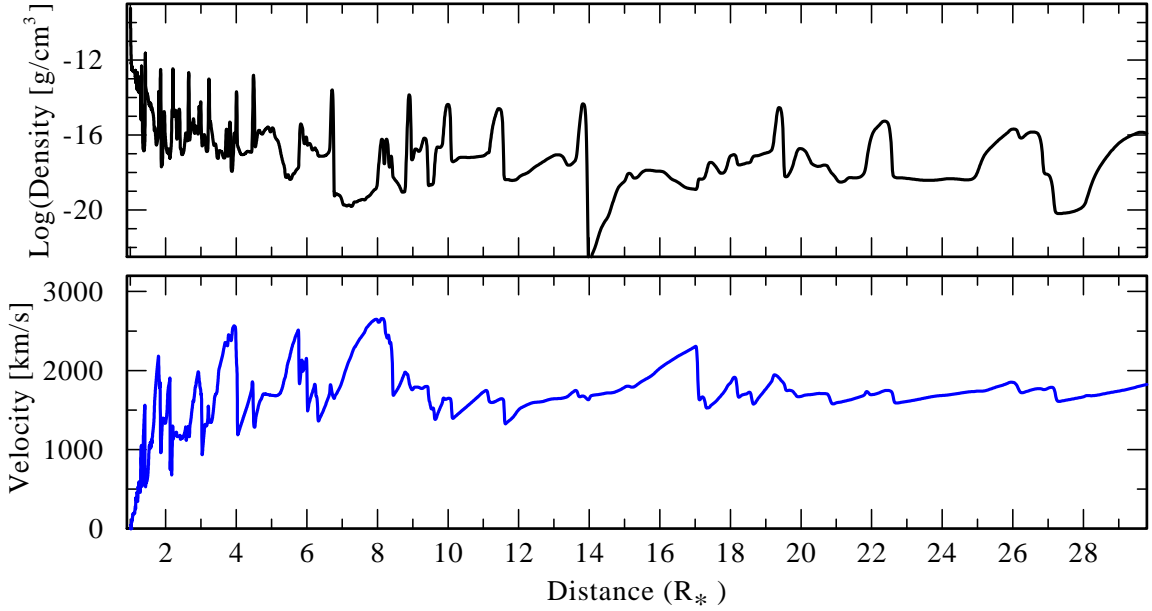}
\caption{A snapshot at a fixed time of the density stratification
(upper panel) and velocity field (lower panel) of the stellar wind in
a O9.5 supergiant star, as predicted by time-dependent hydrodynamic
simulations described in Sect.~\ref{sec:clumpywind}. The upper panel
shows that the wind density can change by several orders of magnitude
due to the presence of ``clumps''. The corresponding strong wind
velocity jumps are visible in the lower panel.}
\label{fig:rhov}
\end{figure*}

\section{Clumpy winds}
\label{sec:clumpywind}

To simulate the wind of the OB supergiant star hosted in a SGXB or
SFXT we use the results from hydrodynamic simulations that derive the
dynamics as well as the thermal structure of a line driven stellar
wind from the first principles \citep[see][for more
details]{feldmeier1995,feldmeier97,feldmeier97b}. This is the same
model adopted by \citet{oskinova12} which predicts highly structured
and non-stationary stellar winds with large density and velocity
gradients. In the model, clumps are produced as a consequence of the
line-driven instability \citep[LDI;][]{lucy80}. The unstable growth of this instability is
triggered by seed perturbations at the base of the wind in form of
turbulent variations of velocity and density at a level of roughly one
third of the sound speed. These perturbations have a coherence time of
1.4\,hr, which is close to the acoustic cutoff period of the model
star. The assumed stellar parameters are typical of a late O-type
supergiant, but it was shown that the predicted stellar wind dynamic
structures of the model are not sensitive to (reasonable) variations
in the input stellar parameters.

Figure\,\ref{fig:rhov} shows a snapshot at a fixed time of the wind
structure produced by the model as a function of the distance from the
massive star. The model predicts a quasi-continuous hierarchy of
density and velocity structures in the wind, where dense shells are
formed relatively close to the star during a first stage of unstable
growth of the LDI. The dense shells have rather small radial extent,
low electron temperatures ($T_{\rm e}\sim$10~kK) and contain the
bulk of the wind mass. The colloquial term ``wind clumps'' refers to
these dense shells. The space between the shells is filled by low
density gas, which is usually termed ``intra-clump'' medium.

Besides the dense cool shells and the tenuous intra-clump medium,
there is a third type of structure. These structures can be described
as small ``clouds'' that are accelerated by the stellar radiation
field and eventually collide with the next-outer dense shell. The
collisions between clouds and shell are the mechanism through which
the model produces the X-ray emission with properties close to those
observed from early type stars at high energies
\citep{feldmeier97,feldmeier97b,Oskinova2006}.

In these time-dependent hydrodynamic simulations, the time-averaged
velocity field of the stellar wind follows the so-called $\beta$-law,
$v_{\rm w}(r)=v_\infty(1-1/r)^\beta$, where $v_\infty$ is the wind
terminal speed and $r$ is the distance from the center of the massive
star in units of the stellar radius. Such velocity field is naturally 
obtained as an outcome of the hydrodynamical model. Strong velocity jumps are
produced in the model, with negative gradients across the shells. Even
relatively close to the star, the wind velocity can change by hundreds
of km/s within a few hours. The model includes the incremental growth
of the optical depth due to multiple resonances in a non-monotonic
velocity field, and is able to resolve the cooling zones behind strong
shocks. This provides a complete and detailed hydrodynamic description
of the stellar wind (radiation, temperature, density, and velocity),
which is currently possible only in similar 1D models.

\section{Gated accretion models}
\label{sec:gating}

In this section we review the gating accretion model 
presented by B08 that illustrates all the different accretion regimes that a NS hosted 
in a SGXB or an SFXT can experience \citep[see also][for previous treatments of the wind  
accretion regimes]{lipunov87,lipunov92}. 
At odds with the more simplified approach of the Bondi-Hoyle accretion 
(see Appendix.~\ref{sec:appendix1}) adopted by \citet{oskinova2012}, the gating accretion model 
takes into account the effect produced onto the accretion flow by the NS spin and magnetic field. 
The characteristics and the main physical processes dominating the X-ray emission in each accretion regime 
are briefly summarized below. We refer the reader to the original paper of B08 for further details 
on the model and on all involved assumptions.   
We first define: 
\begin{itemize}

\item The accretion radius, R$_{\rm a}$, that is the distance from the NS at which the 
inflowing stellar wind material is gravitationally focused toward the compact object: 
\begin{equation}
R_{\rm a}=2GM_{\rm NS}/v_{\rm rel}^2 \sim 3.7\times10^{10} v_{8}^{-2} ~{\rm cm},
\label{eq:ra} 
\end{equation}
where $v_{\rm rel}$ is the relative velocity of the NS compared to the stellar wind and 
v$_{8}$ is the stellar wind velocity in units of 1000~km~s$^{-1}$. It 
has been assumed here for simplicity that the orbital velocity of the NS  
is negligible compared to $v_{\rm w}$ (as done also in B08). 
 
\item The magnetospheric radius, $R_{\rm M}$, at which the pressure of the NS magnetic 
field balances the ram pressure of the inflowing matter. If the magnetospheric radius is larger than 
the accretion radius R$_{\rm a}$, the former can be calculated as: 
\begin{equation}
R_{\rm M} = R_{\rm M1}=1.3\times10^{10} \rho_{-12}^{-1/6} v_{8}^{-1/3} \mu_{33}^{1/3} ~ {\rm cm}.
\label{eq:rm1} 
\end{equation} 
Here\footnote{We approximated a-R$_{\rm M}$$\simeq$a,  
which is satisfied for a very wide range of parameters of interest for the HMXBs considered in this paper.} 
$\mu_{33}$ = $\mu$/10$^{33}$~G~cm$^{3}$ is the NS dipolar magnetic field and 
$\rho_{-12}$=$\rho_{\rm w}$/(10$^{-12}$ g~cm$^{-3}$) is the density of the stellar wind close to the 
compact object. 

\item The corotation radius, R$_{\rm co}$, at which the velocity of the NS spin rotation  
equals the local Keplerian angular velocity, i.e. 
\begin{equation}
R_{\rm co}=1.7\times10^{10} P_{\rm s3}^{2/3} ~{\rm cm},    
\label{eq:rco} 
\end{equation}
where $P_{\rm s3}$ is the NS spin period in units of 10$^3$~s. 
\end{itemize}

We additionally define the orbital separation between the NS and the supergiant companion as 
$a$ = 4.2$\times$10$^{12}$a$_{\rm 10d}$~cm, where 
a$_{\rm 10d}$=P$_{\rm 10d}^{2/3}$M$_{30}^{1/3}$, P$_{\rm 10d}$ is the binary orbital 
period in units of 10 days, and M$_{30}$ is the total 
mass of the system (NS + supergiant companion) in units of 30~M$_{\odot}$. 
We assume in all cases a circular orbit, as done originally in B08. Note that all 
radii defined above are much smaller than the supergiant radius and the radial extent 
of the clump structures mentioned in Sect.~\ref{sec:clumpywind}, thus justifying the usage 
of the 1D model for these systems. 

The conditions under which the different accretion regimes can set-in depend mostly on the relative position of the accretion 
radius, the corotational radius, and the mangetospheric radius. As shown by the equations above, 
the corotation radius is only a function of the NS spin period and can thus be considered constant 
once $P_{\rm spin}$ is fixed\footnote{In this paper we will only consider accretion processes over relatively 
short timescales (hours to days) compared to those on which we expect the NS spin period to change significantly 
as a consequence of accretion torques (see B08).}. The accretion radius and the magnetospheric radius 
are, instead, strongly dependent from the properties of the stellar wind and can thus change significantly 
if the properties of the medium surrounding the compact object are altered, e.g., by the presence of a clump. 
We have to distinguish a number of different possibilities, as illustrated in the following sections. 
\begin{figure}
\centering
\includegraphics[scale=0.6]{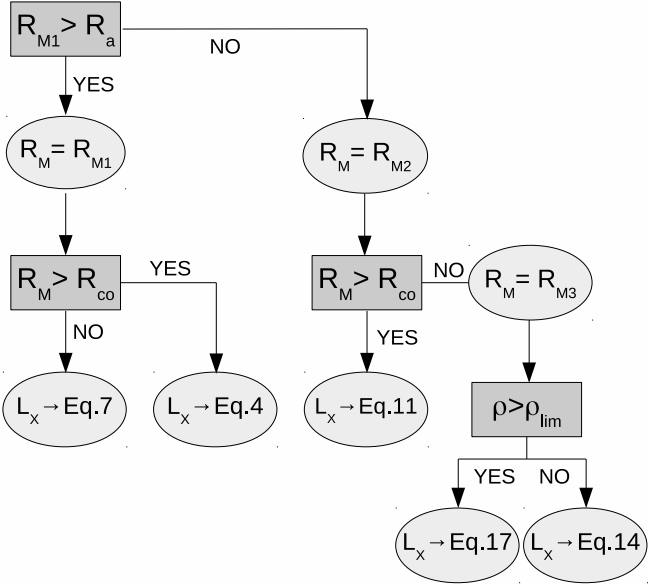}
\caption{Schematic flow of all different conditions that are considered within the clumpy wind code 
to establish which is the relevant accretion regime for the system at each time step and how the total 
X-ray luminosity should be calculated.} 
\label{fig:diagram}
\end{figure}
\begin{figure*}
\centering
\includegraphics[scale=0.42]{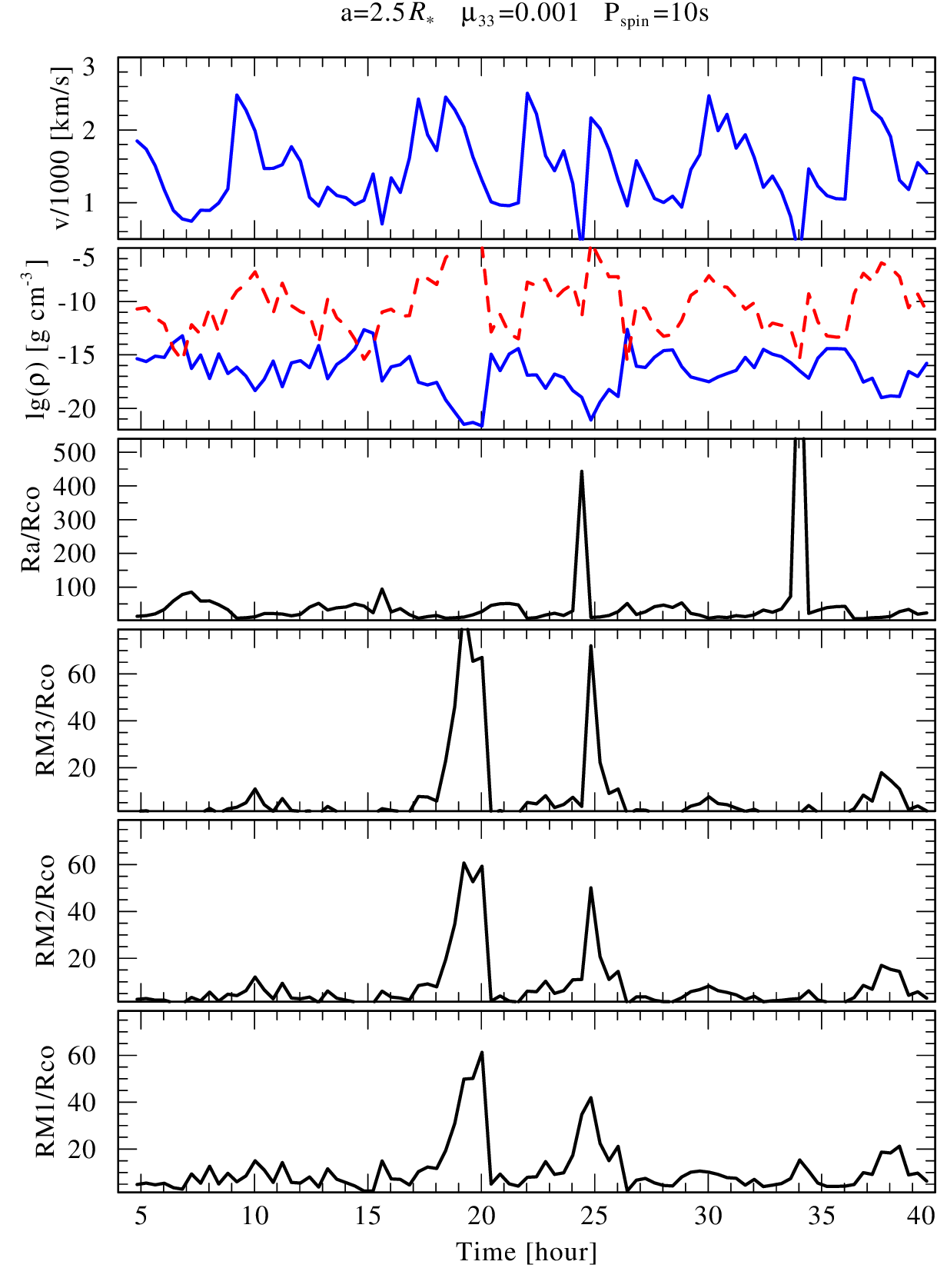}
\hspace{0.5cm}
\includegraphics[scale=0.42]{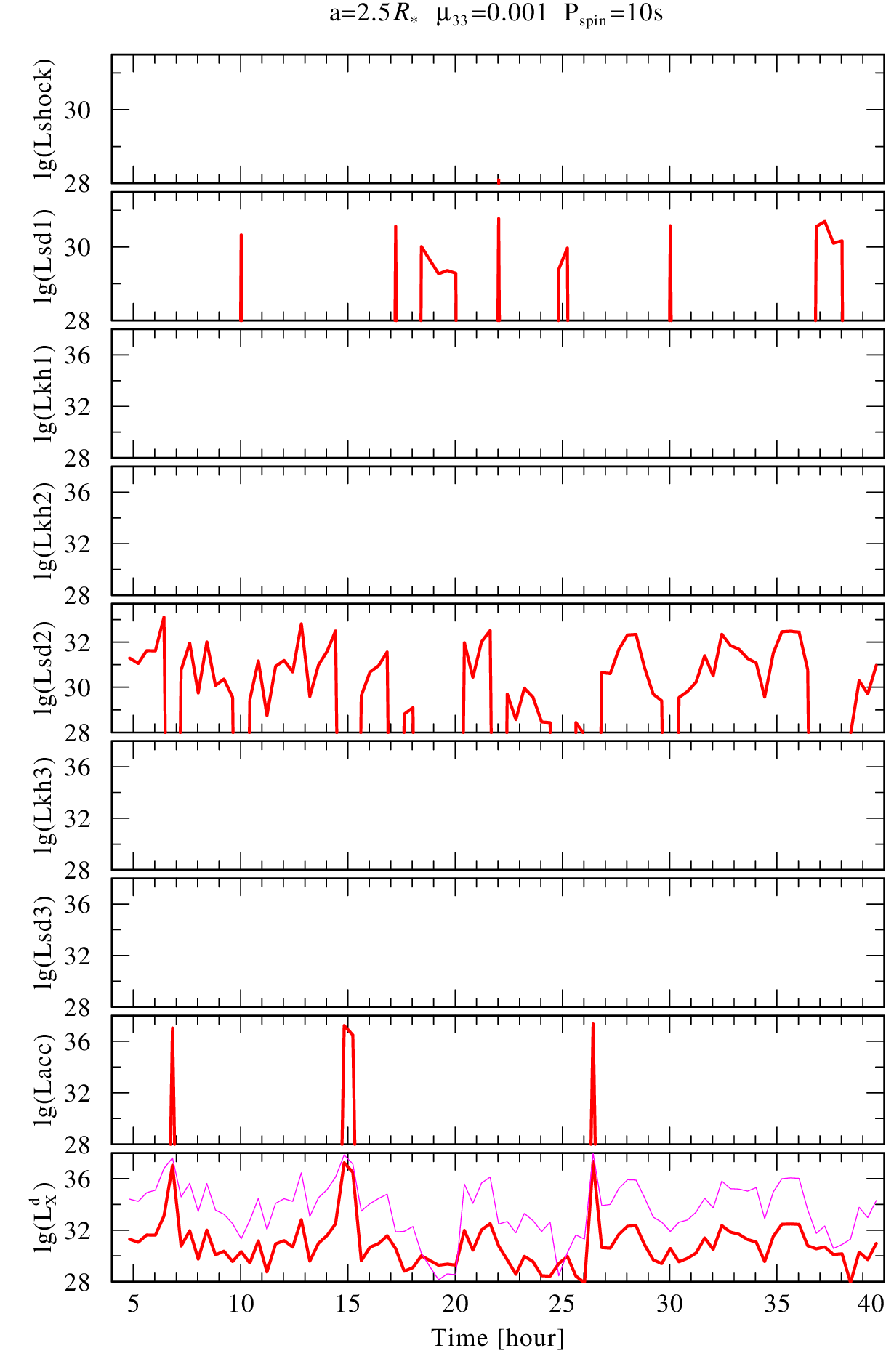}
\caption{Results of the simulations of the accretion onto a NS using the non-stationary wind model introduced 
in Sect.~\ref{sec:clumpywind} and taking into account the gating accretion mechanisms described in Sect.~\ref{sec:gating}. 
The system parameters adopted in the simulation are shown on the top of each figure 
(note that a separation of 2.5~R$_{*}$ corresponds to an orbital period of 9.1~days). The left figure reports the wind velocity and 
density as a function of time, and all relevant radii to be determined in the gating accretion model. 
In the density panel we also show with a red dashed line the value of the critical density 
in Eq.~\ref{eq:rholim}. The right figure shows in all its panels the different contributions to the total X-ray 
luminosity of the system. The latter is reported in the lowermost panel, together with the corresponding 
X-ray luminosity that would be achieved by assuming the simplest Bondi-Hoyle scenario (magenta solid line; 
see Appendix~\ref{sec:appendix1}). Empty regions in each panel represent  
time intervals in which the corresponding quantity is not calculated as the source is not in that specific 
accretion regime (or it is out of scale and thus negligible for the overall luminosity budget; see Sect.~\ref{sec:results} 
for details).} 
\label{fig:short1}
\end{figure*}

\subsection{R$_{\rm M1}$$>$R$_{\rm a}$}

We begin by considering the case in which the magnetospheric radius is larger than the accretion radius. 
Before entering the details of the X-ray emission released in this regime we further have to distinguish between the 
two cases below: 
 
\begin{itemize}
\item R$_{\rm M1}$$\geq$R$_{\rm co}$. If R$_{\rm M1}$$>$R$_{\rm a}$ and R$_{\rm M1}$$>$R$_{\rm co}$, we are in the 
so called super-Keplerian magnetic inhibition regime. In this case, both the magnetic and the centrifugal 
gates are closed and inhibit the accretion onto the NS. In particular, the magnetic gate prevents matter for being 
gravitationally focused toward the NS and the centrifugal gate propels away the material along the magnetospheric boundary of the 
compact object. The X-ray luminosity of the system is dominated in this case by 
the energy released in the shocks occurring close the magnetospheric boundary (R$_{\rm M}$$=$R$_{\rm M1}$) and 
by friction between the rotating NS magnetosphere and the surrounding material. We thus have:  
\begin{equation}
L_{X}=L_{\rm shock}+L_{\rm sd1}\,,
\end{equation}
where
\begin{equation}
L_{\rm shock}\simeq\frac{\pi}{2} R_{\rm M1}^2 \rho_{\rm w} v_{\rm w}^3 = 
2.7\times10^{32} \mu_{33}^{2/3} \rho_{-12}^{2/3} v_{8}^{7/3} ~{\rm erg ~s^{-1}}  
\label{eq:lxshock1}
\end{equation}
and 
\begin{eqnarray}
L_{\rm sd1} \simeq \pi R_{\rm M}^2 \rho_{\rm w} v_{\rm w} (R_{\rm M1}\Omega)^2 \simeq  
3.5\times10^{32} \mu_{33}^{4/3} \cdot \nonumber \\ \rho_{-12}^{1/3} v_{8}^{-1/3} P_{\rm s3}^{-2}~{\rm erg ~s^{-1}}. 
\label{eq:lxsd1}
\end{eqnarray}
 
\item R$_{\rm M1}$$<$R$_{\rm co}$. In this case only the magnetic gate is closed and we are in the sub- Keplerian magnetic 
inhibition of accretion. Matter is still not gravitationally focused toward the compact object but since the propeller effect is not 
effective, the material passing along the magnetospheric boundary of the compact object is not pushed away and can accrete mainly 
through the Kelvin-Helmholtz instability (hereafter KHI). As shown by B08, the total X-ray luminosity is given in this case by: 
\begin{equation}
L_X=max(L_{\rm KH1},L_{\rm KH2})\,,
\label{eq:lkh12}
\end{equation}
where 
\begin{eqnarray}
L_{\rm KH1} = 2.0\times10^{37} \eta_{\rm KH} \mu_{33}^{2/3}\rho_{-12}^{2/3} v_{8}^{1/3}
(\rho_{\rm i}/\rho_{\rm e})^{1/2} \cdot  \nonumber \\ (1+\rho_{\rm i}/\rho_{\rm e})^{-1}~{\rm erg~s^{-1}}, 
\end{eqnarray}
and 
\begin{eqnarray}
L_{\rm KH2} = 6.5\times10^{37} \eta_{\rm KH} P_{\rm s3}^{-1} v_8^{-1} \rho_{-12}^{1/2} \mu_{33}  
(\rho_{\rm i}/\rho_{\rm e})^{1/2} \cdot  \nonumber \\ (1+\rho_{\rm i}/\rho_{\rm e})^{-1} ~{\rm erg~s^{-1}}.  
\label{eq:lx_kh2}
\end{eqnarray}
We assume for the present work the reasonable value of 0.5 for the ratio between the wind density material inside ($\rho_{\rm i}$) 
and outside ($\rho_{\rm e}$) the compact object mangetospheric boundary. We also use a standard value for the 
parameter of the KHI efficiency $\eta_{\rm KH}=0.1$ (B08). Higher values of this 
quantity will only slightly increase the X-ray luminosity in the sub-Keplerian magnetic inhibition regime 
but do not qualitatively change the results discussed in Sect.~\ref{sec:results}. 
\end{itemize}
\begin{figure}[ht!]
\centering
\includegraphics[scale=0.35]{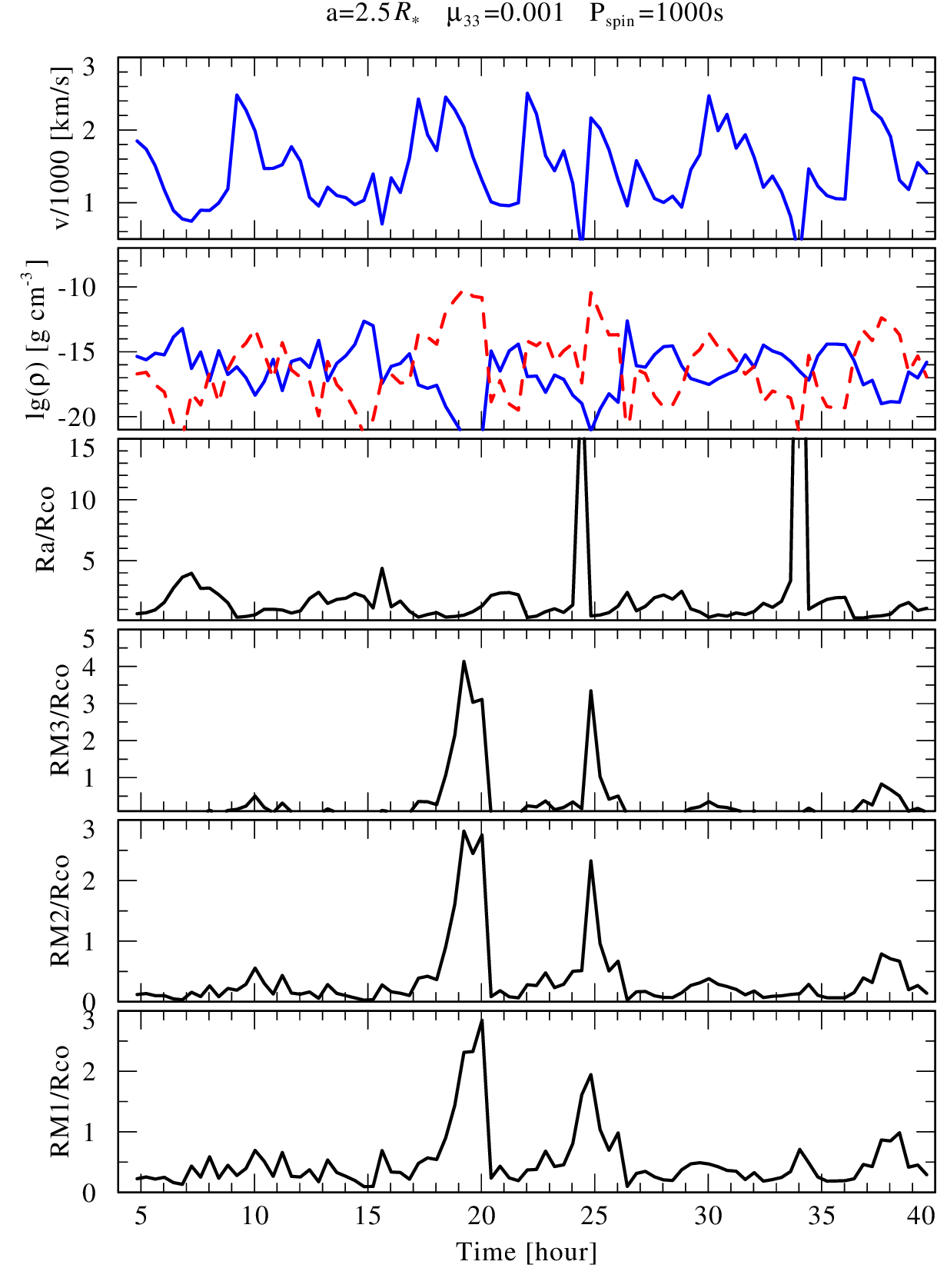}
\includegraphics[scale=0.35]{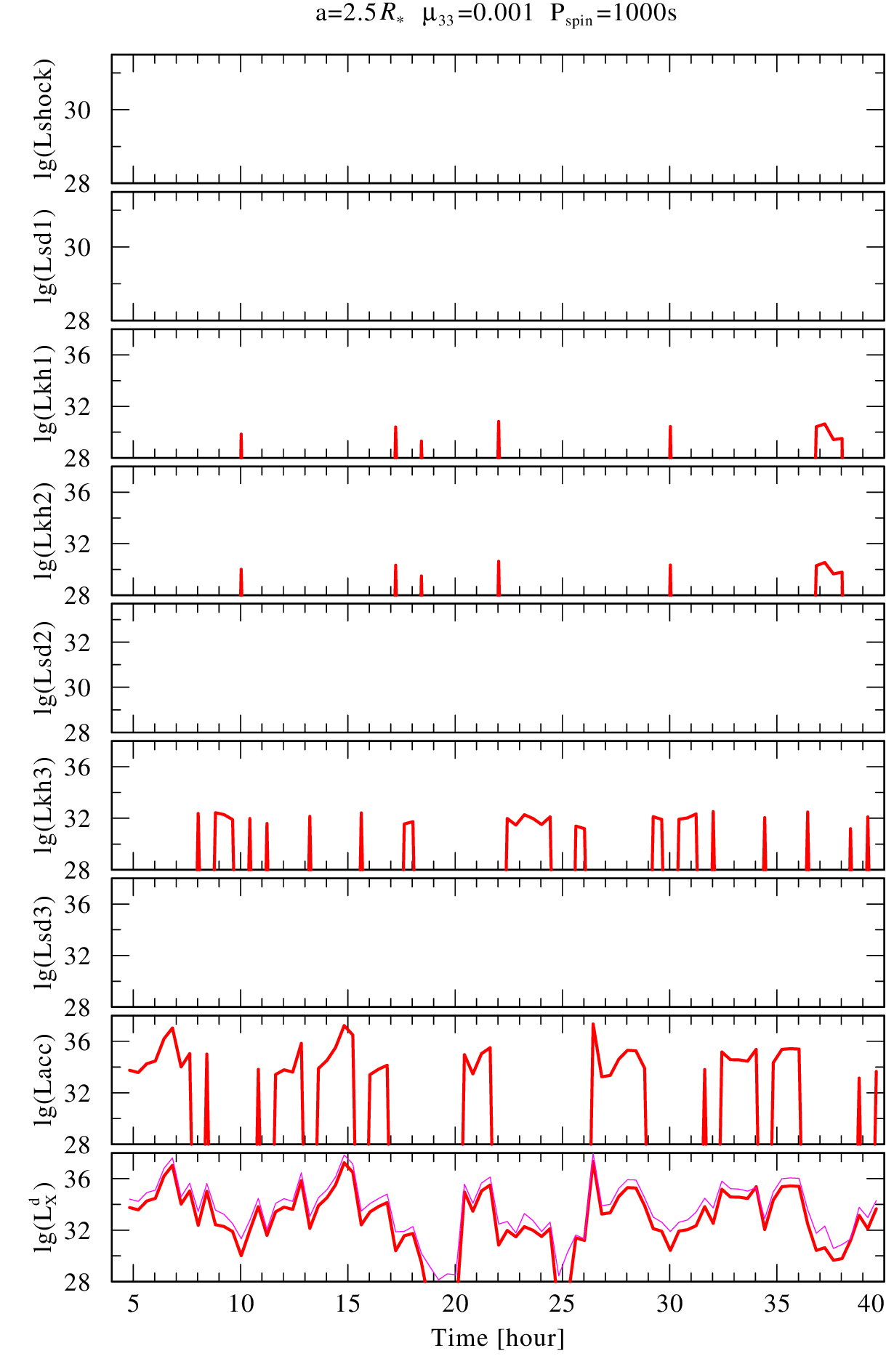}
\caption{Same as Fig.~\ref{fig:short1} but for a NS spin period of 1000~s. 
The magnetic field strength was not changed.} 
\label{fig:short2}
\end{figure}
\begin{figure}[ht!]
\centering
\includegraphics[scale=0.35]{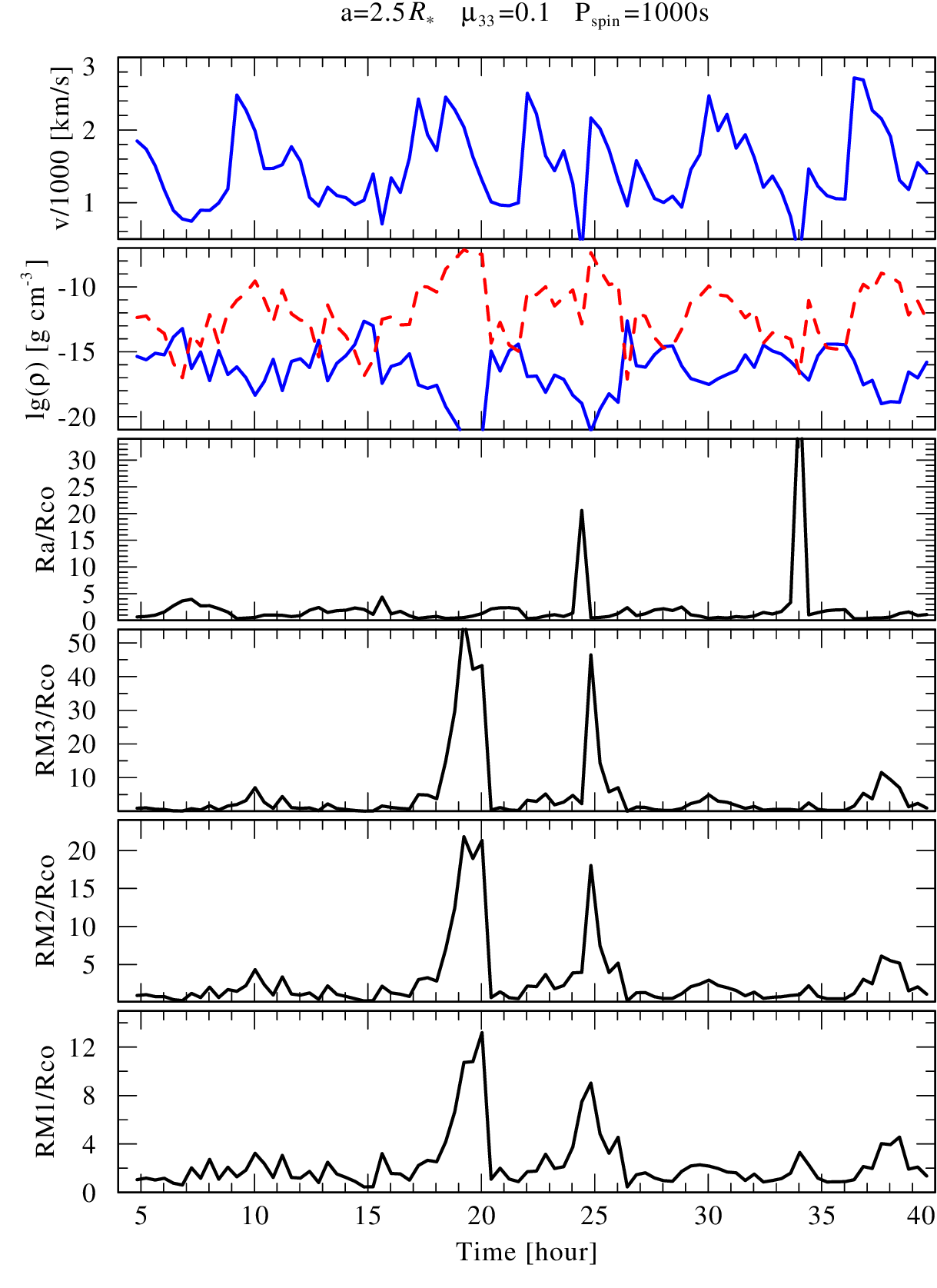}
\includegraphics[scale=0.35]{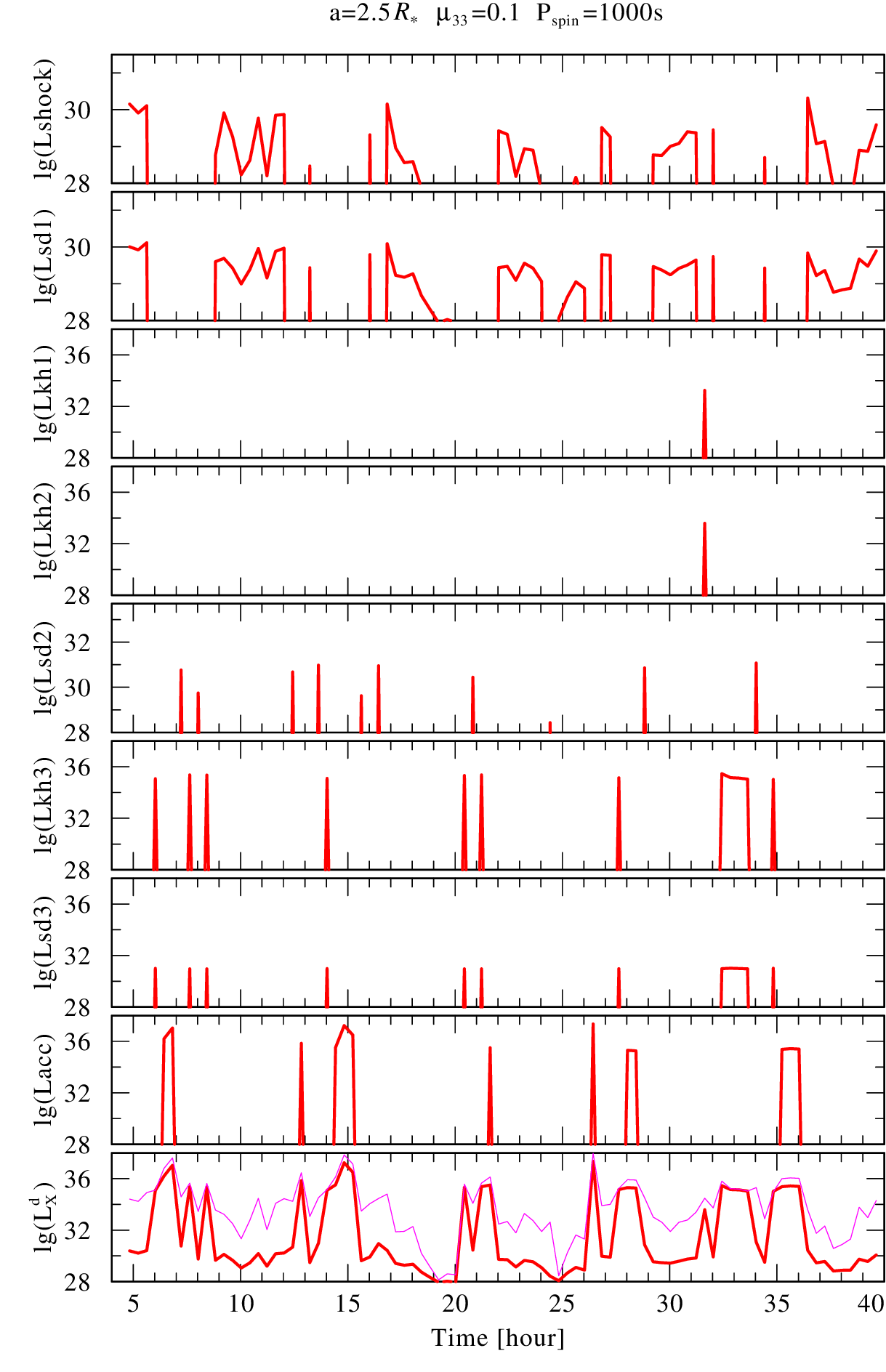}
\caption{Same as Fig.~\ref{fig:short1} but for a NS spin period of 1000~s and a magnetic field 
strength of 10$^{14}$~G.} 
\label{fig:short3}
\end{figure}
\begin{figure}[ht!]
\centering
\includegraphics[scale=0.35]{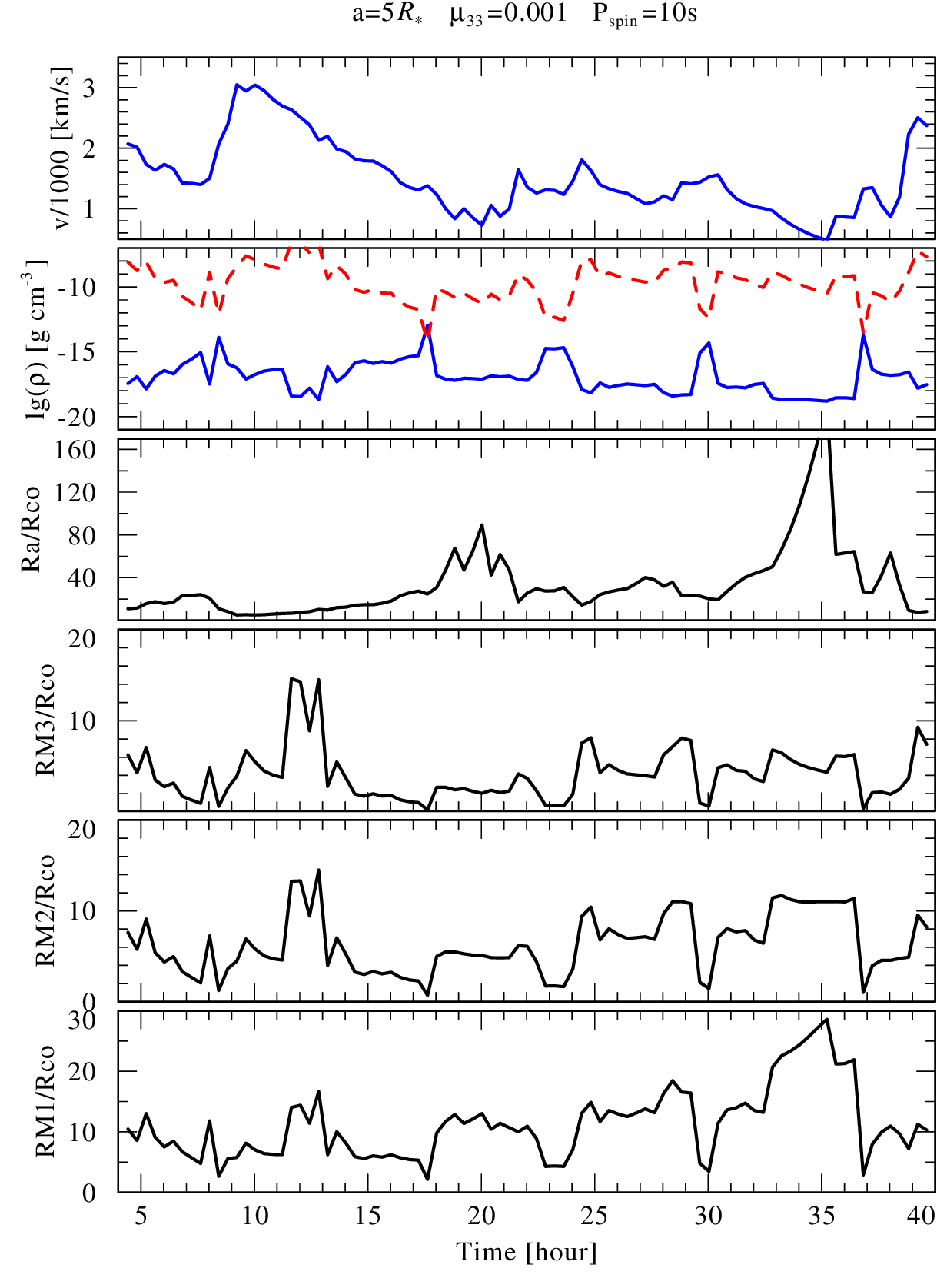}
\includegraphics[scale=0.35]{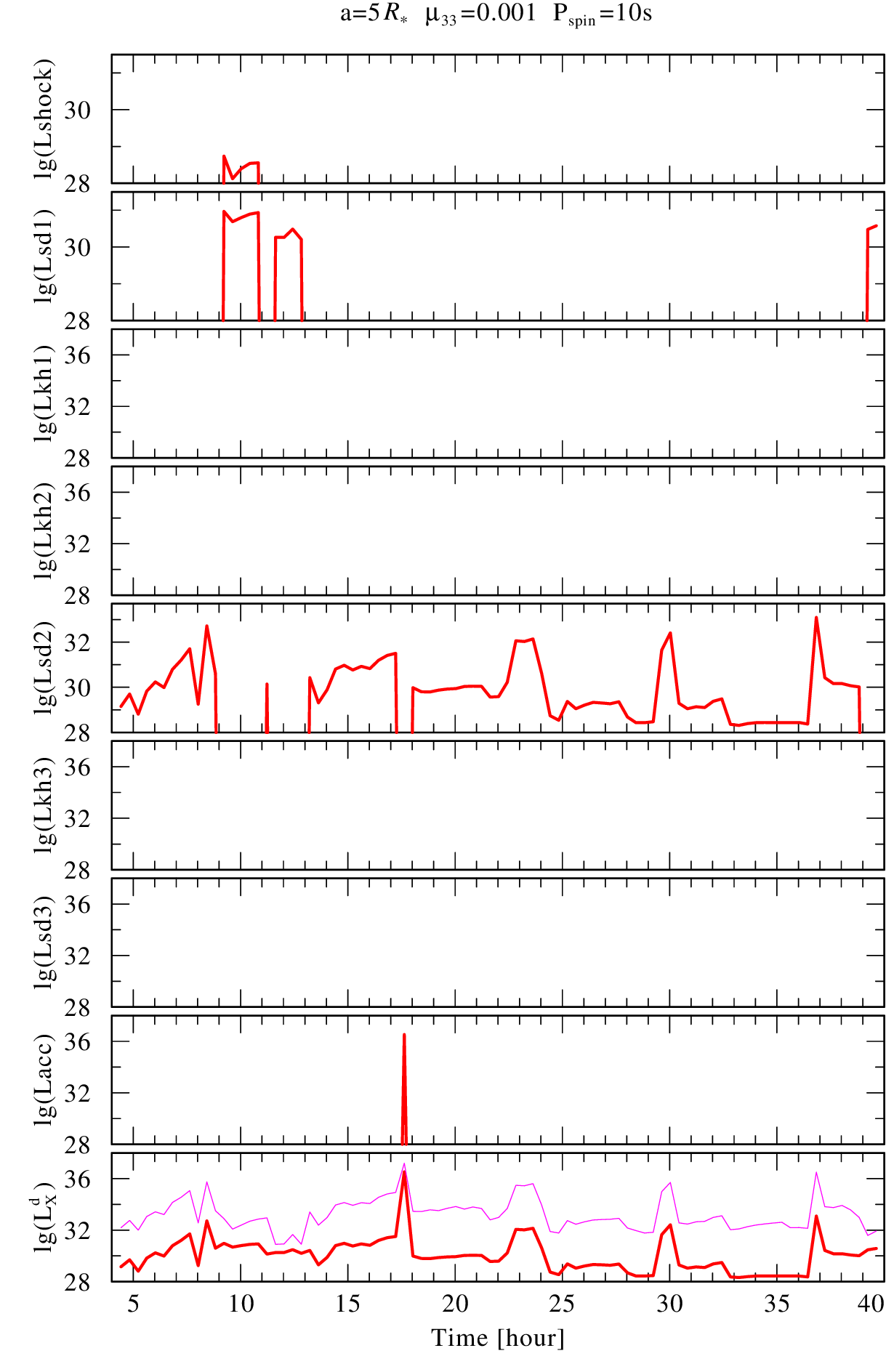}
\caption{Same as Fig.~\ref{fig:short1} but in the case of a larger orbital 
separation between the NS and the supergiant companion. In this case the separation 
corresponds to 25.6~days, assuming a NS of 1.4~$M_{\odot}$ and a supergiant of 
34~$M_{\odot}$.} 
\label{fig:long1}
\end{figure}
\begin{figure}[ht!]
\centering
\includegraphics[scale=0.35]{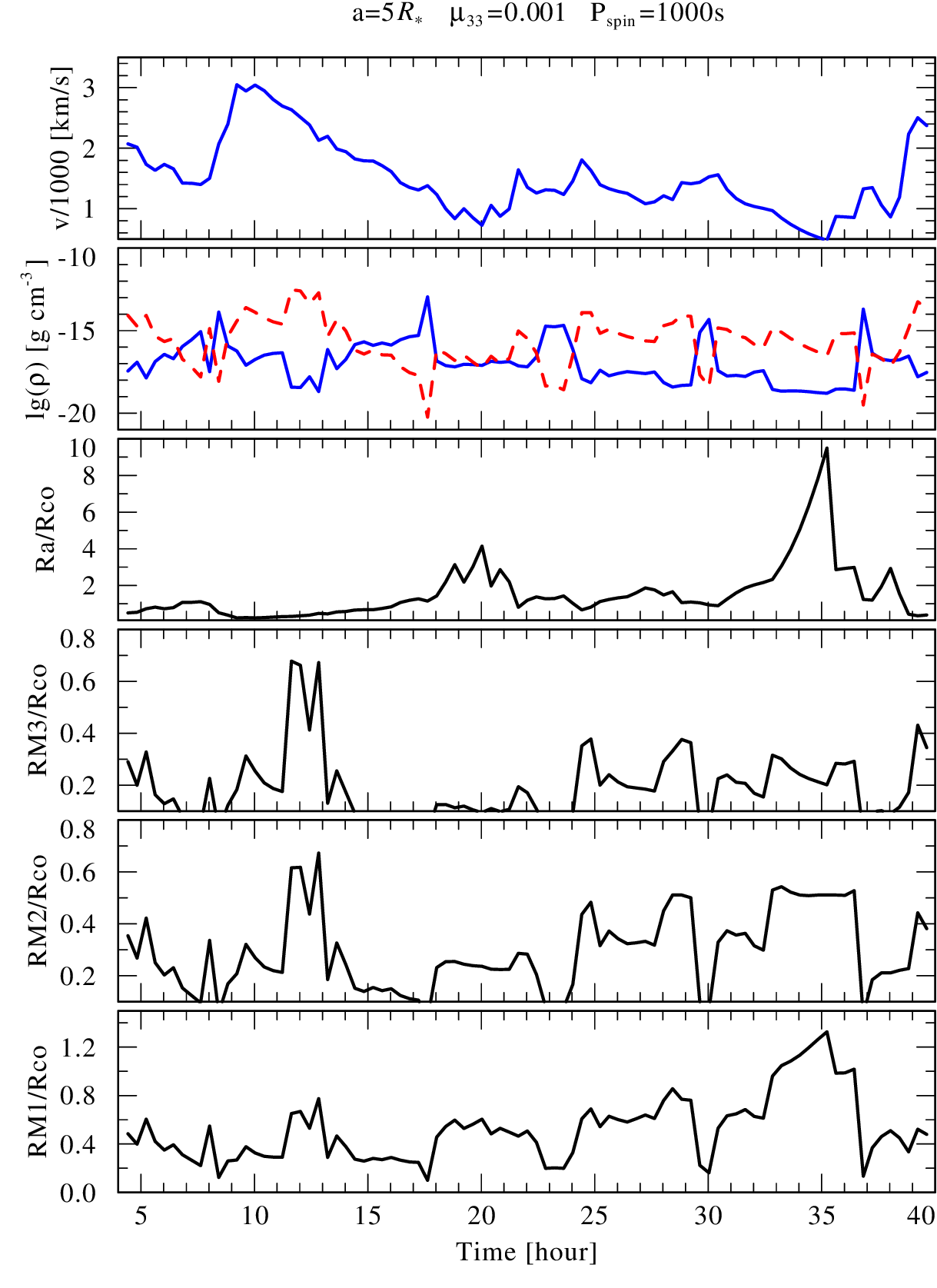}
\includegraphics[scale=0.35]{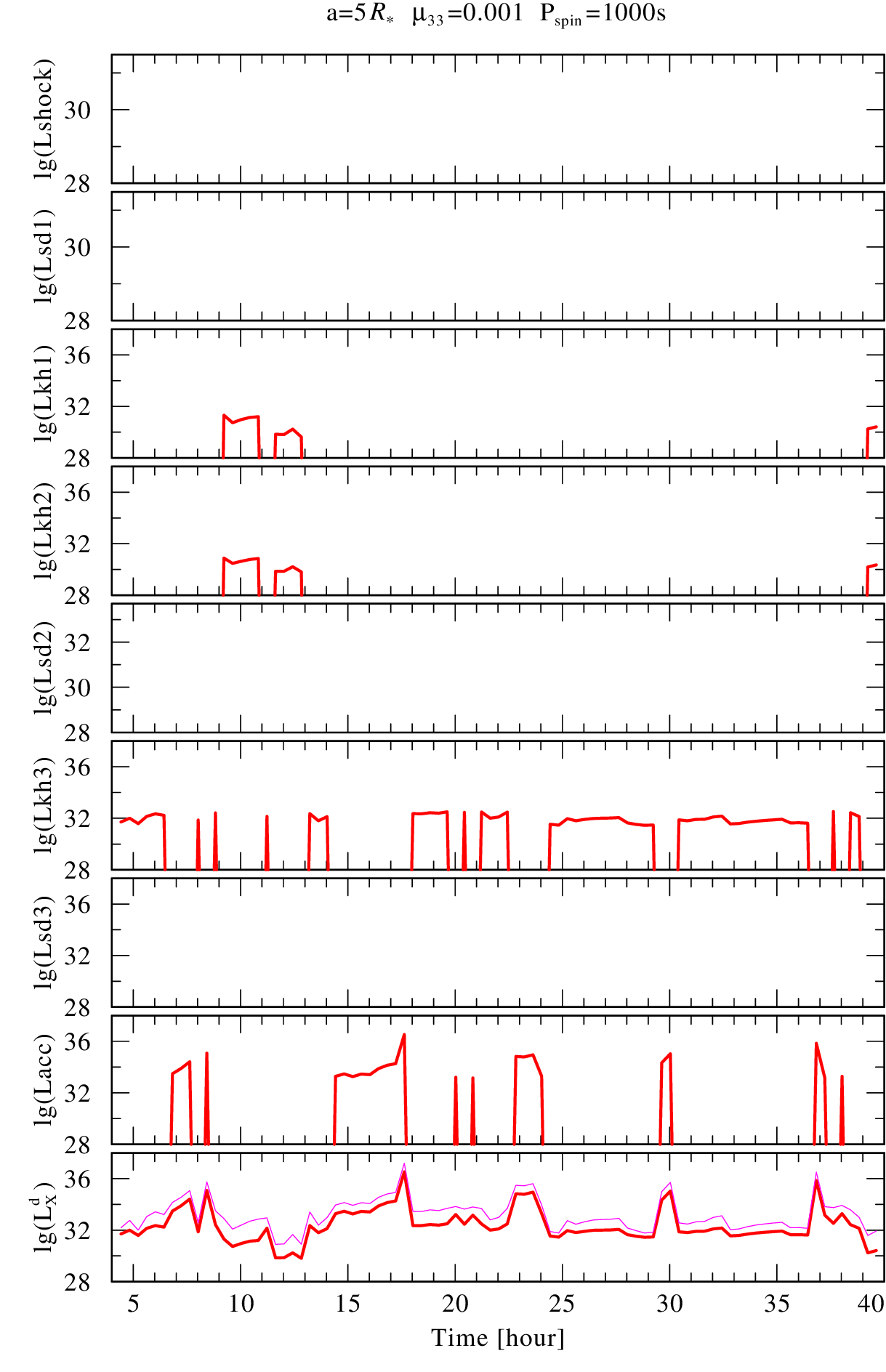}
\caption{Same as Fig.~\ref{fig:long1} but for a NS spin period of 1000~s. 
The magnetic field strength was not changed.} 
\label{fig:long2}
\end{figure}
\begin{figure}[ht!]
\centering
\includegraphics[scale=0.35]{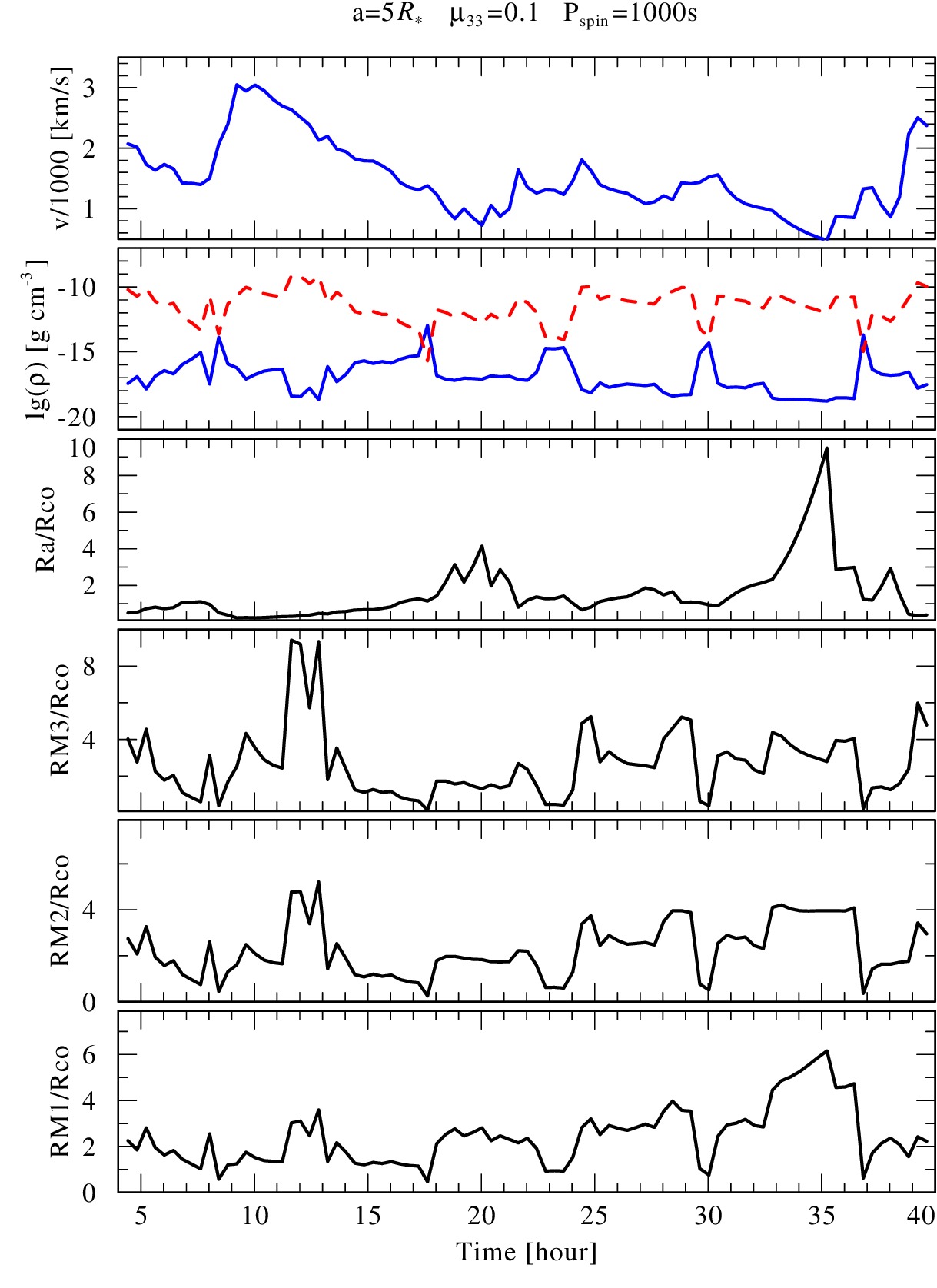}
\includegraphics[scale=0.35]{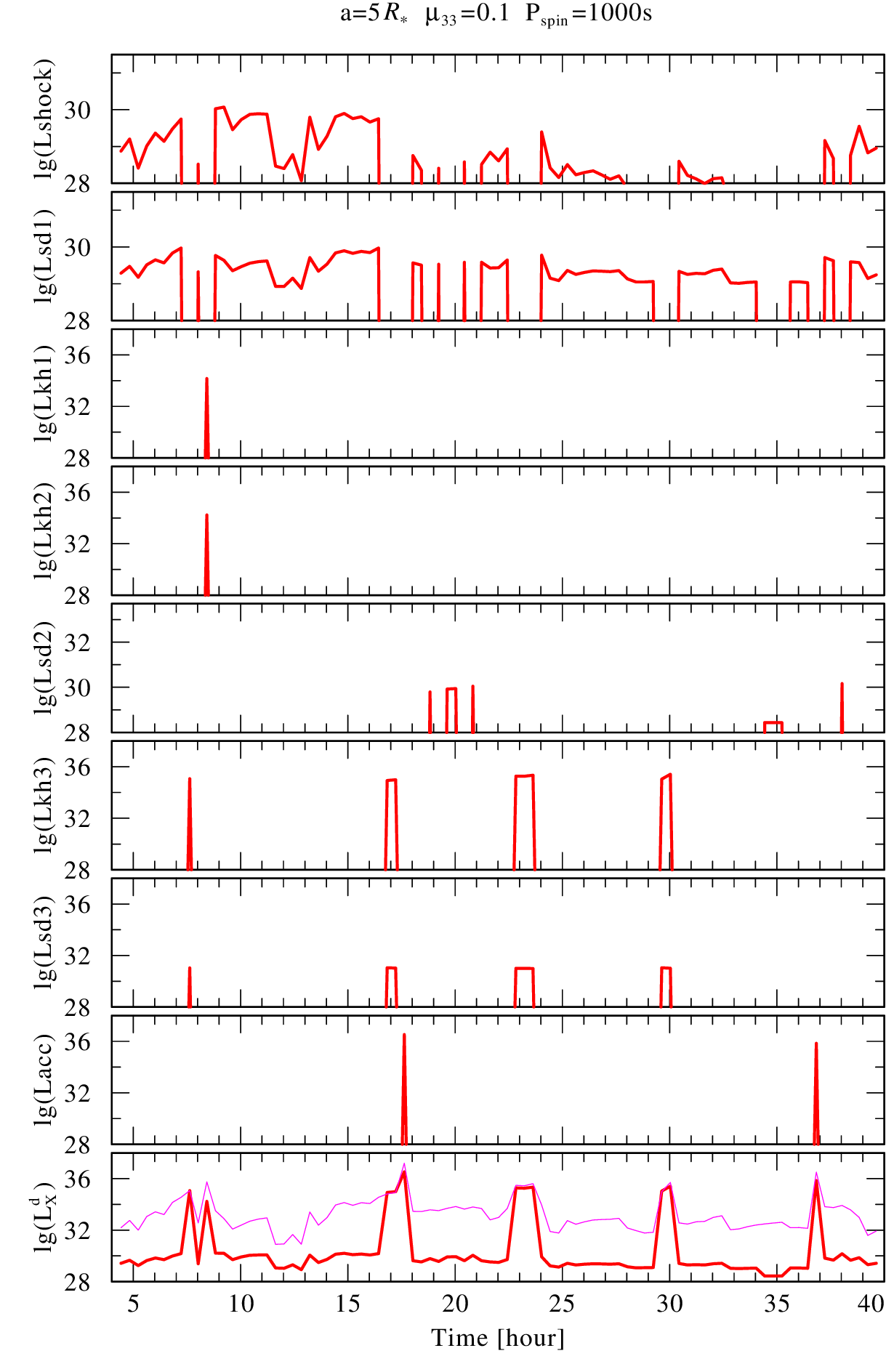}
\caption{Same as Fig.~\ref{fig:long1} but for a NS spin period of 1000~s and a magnetic field 
strength of 10$^{14}$~G.} 
\label{fig:long3}
\end{figure}

\subsection{R$_{\rm M1}$$<$R$_{\rm a}$}

When the magnetospheric radius R$_{\rm M1}$$<$R$_{\rm a}$, the magnetic gate is open and the inflowing material from the 
stellar wind can be gravitationally focused toward the compact object, filling the space between R$_{\rm a}$ and the magnetospheric radius.  
This material redistributes itself into an approximately spherical configuration, resembling an ‘‘atmosphere’’, whose shape and 
properties are determined by the interaction with the rotating NS magnetosphere at the magnetospheric boundary.

The equation defining the magnetospheric radius changes as the pressure of the material is now computed 
assuming an hydrostatic equilibrium and we have R$_{\rm M}$$=$R$_{\rm M2}$, where: 
\begin{equation}
R_{\rm M2}=6.8\times10^{9} v_8^{2/9} \mu_{33}^{4/9} \rho_{-12}^{-2/9} ~{\rm cm} .
\label{eq:rm2} 
\end{equation}

Before computing the X-ray luminosity released in this case, we have to distinguish between the different cases below: 

\begin{itemize}

\item R$_{\rm M2}$$>$R$_{\rm co}$: in this case the centrifugal gate is closed and the rapid rotation of the NS 
propels away the inflowing material after it has been gravitationally focused. We are thus in the so-called supersonic propeller regime. 
Matter entered within the accretion radius cannot accrete onto the NS and the main contribution to the X-ray luminosity 
of the system is due to the friction of the material around the NS and the compact object magnetosphere:  
\begin{eqnarray}
L_{\rm X} = L_{\rm sd2}= 2\pi R_{\rm M}^2 \rho(R_{\rm M}) c_{\rm s}^3
(R_{\rm M})\simeq  \nonumber \\ 8.2\times10^{34} v_8^{-1} \rho_{-12} ~{\rm erg ~s}^{-1} 
\end{eqnarray}
In the above equation c$_{\rm s}$(R$_{\rm M}$)=v$_{\rm ff}$ 
(R$_{\rm M}$)=(2GM$_{\rm NS}$/R$_{\rm M}$)$^{1/2}$ is the sound velocity in the shell of material 
surrounding the NS.  

\item R$_{\rm M2}$$<$R$_{\rm co}$: if R$_{\rm M2}$ is no longer larger than the corotation radius, 
then both the magnetic and centrifugal gates are open and the inflowing stellar wind material 
can begin penetrating the NS magnetosphere and accretes onto the compact object. 
In this situation, the rotational velocity of the NS magnetosphere is sub-sonic and the magnetospheric 
radius has to be calculated according to the equation below: 
\begin{equation}
R_{\rm M} = R_{\rm M3} = 3.8\times10^{9} v_8^{6/7} \mu_{33}^{4/7} \rho_{-12}^{-2/7}~{\rm cm}. 
\label{eq:rm3} 
\end{equation}

Accretion at full regime can however take place only when there is a sufficiently high density outside $R_{\rm M3}$, such 
that the inflowing wind material can cool down rapidly and the magneto-hydrodynamic instabilities at the NS magnetospheric boundary 
reach the highest efficiency in transporting material inside the NS magnetosphere. 
The critical density above which accretion occurs at full regime (i.e. comparable with the accretion rate in the Bondi-Hoyle approximation) 
is given by:   
\begin{equation}
\rho_{\rm lim_{-12}} = 0.83 P_{\rm s3}^{-3} R_{\rm M10}^{5/2} (1+16 R_{\rm a10}/(5 R_{\rm M10}))^{-3/2}, 
\label{eq:rholim}
\end{equation}
where $R_{\rm M10}$=$R_{\rm M3}$/10$^{10}$. 

We thus still have to distinguish among two additional cases: 

\begin{itemize}

\item If $\rho_{-12}$$<$$\rho_{\rm lim_{-12}}$, then accretion does not achieve the 
highest efficiency and the system enters the subsonic propeller regime. It can be shown that the main contributions to the total released 
X-ray luminosity is provided by the partly limited accretion allowed by the KHI and the friction between the material surrounding the 
NS and its magnetosphere: 
\begin{equation}
L_X=L_{\rm KH3}+L_{\rm sd3}\,,
\end{equation}
where: 
\begin{eqnarray}
& & L_{\rm KH3} \simeq G M_{\rm NS} \dot{M}_{\rm KH}/R_{\rm NS}= 1.5\times10^{38} \eta_{\rm KH} P_{\rm s3}^{-1} R_{\rm M10}^3 \cdot \nonumber \\ 
& & \cdot \rho_{-12} (1+16 R_{\rm a10}/(5 R_{\rm M10}))^{3/2} (\rho_{\rm i}/\rho_{\rm e})^{1/2} \cdot \nonumber \\ 
& & \cdot (1+\rho_{\rm i}/\rho_{\rm e})^{-1} ~{\rm erg ~s}^{-1} 
\label{eq:lkh3}
\end{eqnarray}
and 
\begin{eqnarray}
L_{\rm sd3} =  2\pi R_{\rm M}^5 \rho(R_{\rm M}) \Omega^3 
= 6.2\times10^{32}  P_{\rm s3}^{-3} R_{\rm M10}^5 \rho_{-12} \cdot \nonumber \\ 
(1+16 R_{\rm a10}/(5 R_{\rm M10}))^{3/2} ~{\rm erg ~s}^{-1}.  
\label{eq:lsd3} 
\end{eqnarray}

\item If $\rho_{-12}$$\geq$$\rho_{\rm lim_{-12}}$, then all conditions are satisfied for accretion at full regime to take place. 
The system thus achieve the highest X-ray luminosity and enters the so-called direct accretion regime. In this case the mass 
accretion rate is comparable to the Bondi-Hoyle approximation and we have: 
\begin{eqnarray}
L_{\rm X} = L_{\rm acc}= G M_{\rm NS} \dot{M}_{\rm capt} / R_{\rm NS} = \nonumber \\
5.9\times10^{36} v_8 \rho_{-12} R_{\rm a10}^2 ~{\rm erg ~s}^{-1}. 
\label{eq:lxacc} 
\end{eqnarray}

\end{itemize}

\end{itemize}

All equations in this section have been re-adapted from B08 to be used within the clumpy wind code described in Sect.~\ref{sec:clumpywind}. 
This code provides the value of the wind velocity $v_8$ and the density 
$\rho_{-12}$  as a function of time at the NS location (which is a fixed parameter). These values 
are first used to determine which accretion regime the system is experiencing at a certain time and then estimate the corresponding 
X-ray luminosity. A schematic version of the flow of conditions that lead the code to establish the relevant accretion regime of the system 
and compute the X-ray luminosity at each time step is shown in Fig.~\ref{fig:diagram}.

\section{Results}
\label{sec:results}

We present in this section the results obtained by simulating the accretion of the clumpy wind model 
in Sect.~\ref{sec:clumpywind} onto a strongly magnetized rotating NS.  
The main parameter to be fixed for the clumpy wind model 
is the distance between the compact object and the massive companion, i.e. the orbital separation. As we discussed in 
Sect.~\ref{sec:clumpywind} the properties of the stellar wind change at different distances from the massive stars, thus 
affecting the physical properties of the accretion flow that approaches the NS as a function of time. We 
discuss below the two cases in which the NS is located at $a$=$2.5R_{*}$ and $5R_{*}$ from the massive companion, 
(here $R_{*}$ is the supergiant radius). As we considered in all cases a supergiant star with a mass of 34~$M_{\odot}$ and 
a radius of 24~$R_{\odot}$, the above separations correspond to a NS orbital period of 9.1 and 25.6~days, respectively 
 (a circular orbit is assumed in all cases as we cannot simulate in the present set-up the effect of a non-negligible eccentricity). 
In order to probe different combinations of the accretion regimes, we also considered different cases for the NS 
magnetic field and spin period. In particular, we report here on the three cases in which: (i) $P_{\rm s3}=1$ and   
$\mu_{33}=0.1$; (ii) $P_{\rm s3}=1$ and $\mu_{33}=0.001$; (iii) $P_{\rm s3}=0.01$ and $\mu_{33}=0.001$. 
The first of this possibility is more representative of the accretion onto slowly rotating ``magnetars'', i.e. NS endowed 
with particularly strong magnetic fields. The second and the third cases are more representative of the case of a normal 
pulsar in a SGXBs, endowed with a standard magnetic field of $10^{12}$~G and a longer or shorter spin period.

\subsection{Orbital period 9.1~days}
\label{sec:short}

We show the outcomes of the combination of the clumpy wind code with the different accretion regimes for an orbital 
period of 9.1~days in Fig.~\ref{fig:short1}, \ref{fig:short2}, and \ref{fig:short3}.  

The first of these figures show the fastest spin period ($P_{\rm s3}=0.01$) and lowest magnetic field 
($\mu_{33}=0.001$) case considered in this paper. The two uppermost panels of Fig.~\ref{fig:short1} (left) show 
the velocity and density evolution as a function of time computed within the clumpy wind model 
in Sect.~\ref{sec:clumpywind}. In the density panel, we also over-plot with a dashed red line 
the value of the critical density in Eq.~\ref{eq:rholim}. The other panels show from bottom to top the 
different magnetospheric radii $R_{\rm M1}$, $R_{\rm M2}$, and $R_{\rm M3}$, applicable to the different 
accretion regimes and measured in units of the corotation radius $R_{\rm co}$. We also display in the same units 
the accretion radius, $R_{\rm a}$. Depending on the relative position of these radii, it can be understood 
how changes in the local environment surrounding the NS trigger the switch between different accretion regimes 
(see Fig.~\ref{fig:diagram}).  
In the other panels of the figure, we report the values of the luminosity contribution of each accretion regime 
to the total X-ray emission. The latter is reported on the bottom panel (red solid line), together with  
the X-ray luminosity that the system would display in case the simplest Bondi-Hoyle approximation is assumed  
\citep[as done in][see also Appendix~\ref{sec:appendix1}]{oskinova2012}. 
 
For the low magnetic field selected for this first case, the magnetospheric radius is smaller 
than the accretion radius for most of the time and thus the magnetic gate is almost always open (i.e. no inhibition 
of accretion occur due to the magnetic barrier). The only 
exceptions occur when there is an abrupt increase in the velocity of the wind not accompanied by a too large drop in the density 
(see, e.g., $t$=10~h, 17~h, 19~h, 22~h, 25~h, 30~h, 37~h).  
This leads to $R_{\rm M1}\gg R_{\rm a}$ and thus to the closure of the magnetic barrier (note that $R_{\rm a}$ is inversely proportional 
to the wind velocity while $R_{\rm M1}$ is inversely proportional to the wind density). The relatively short spin period of the NS 
makes in all these cases the spin down luminosity given by Eq.~\ref{eq:lxsd1} dominating over the 
contribution due to shocks in the vicinity of the compact object (Eq.~\ref{eq:lxshock1}). 
As the spin period of the NS is relatively small, and so is 
the corotation radius, the system remains in the supersonic propeller regime (i.e., $R_{\rm M1}\ll R_{\rm a}$ and 
$R_{\rm M}=R_{\rm M2}\geq R_{\rm co}$) for a large fraction of the simulated 
time interval. Only occasionally during the entire run (40~hours) the wind density raises above the critical value $\rho_{\rm lim}$ and 
the system switches to the accretion regime (see, e.g., $t$=7~h, 15~h, and 26~h). Overall, Fig.~\ref{fig:short1} 
shows that even if the NS is endowed with a relatively common magnetic field strength and short spin period, the introduction 
of the gating mechanisms lead to a substantial decrease of the average luminosity compared to the simplified Bondi-Hoyle 
accretion scenario (this can be seen by comparing the magenta and red solid lines in the right bottom panel of Fig.~\ref{fig:short1}). 
Although the X-ray variability in this case is remarkably pronounced due to the large variations in the 
wind density/velocity and approaches the dynamic range typical of the SFXTs, the simulated behavior does not closely resemble what 
we observe in these latter sources. In particular, we are missing the presence of the long periods of low level emission that occur 
between the much less frequent outbursts and flares. As noticed already also by \citet{oskinova12}, the observed variability is also  
much larger than what we expect in classical SGXBs.  

In Fig.~\ref{fig:short2} we show the case in which the NS spin period is much longer than before (1000~s) and maintained 
unchanged the value of the magnetic field strength.  
The main difference compared to the results presented in Fig.~\ref{fig:short1} is that the corotation radius of the 
NS is now much larger (as it scales with $P_{\rm s3}^{2/3}$, see Eq.~\ref{eq:rco}) and thus it is more difficult 
for the system to enter the super-Keplerian magnetic inhibition of accretion and/or the supersonic propeller regime. 
As the magnetic field of the NS is still relatively weak ($10^{12}$~G), the system can enter the magnetic inhibition 
of accretion only when the largest drops in the wind density (or increases in the wind velocity) occur, leading to a significant expansion 
of the magnetospheric radius beyond the accretion radius (see Eq.~\ref{eq:rm1} and \ref{eq:ra}). 
Note that also in these cases, the magnetospheric radius remains within the corotation radius, and the dominant 
contribution to the X-ray luminosity of the system is provided by Eq.~\ref{eq:lkh12}. For most of the time, 
the system is in the direct accretion regime, achieving the high X-ray luminosity expressed by Eq.~\ref{eq:lxacc}. 
In this regime the magnetospheric radius is smaller than both the accretion and the corotation radius, and the 
condition on the critical wind density is also satisfied (see Eq.~\ref{eq:rholim}). From Fig.~\ref{fig:short2} 
we note that in several occasion the system also switches to the subsonic propeller regime because, even if the 
magnetospheric radius is smaller than $R_{\rm co}$ and $R_{\rm a}$, the wind density happens to be lower than 
$\rho_{\rm lim}$ at certain intervals of time. In all these cases, the dominant contribution to the system X-ray luminosity is provided 
by Eq.~\ref{eq:lkh3} (note that the contribution given by Eq.~\ref{eq:lsd3} is much lower due to the  
long spin period of the NS assumed in the present case). 

In Fig.~\ref{fig:short3} we increased the strength of the NS magnetic field up to 10$^{14}$~G and kept the previous value 
of the spin period. As it can be seen from this figure, the larger magnetic field pushes the magnetospheric radius beyond the accretion 
radius for a substantial amount of time (note that the time variability of the wind density and velocity is always the same in all 
plots corresponding to the same orbital separation). The system thus spends a significant amount of time in the super-Keplerian magnetic inhibition of accretion, where the 
most relevant contributions to the X-ray luminosity are given by Eq.~\ref{eq:lxshock1} and \ref{eq:lxsd1}. 
Note that, at odds with the case of Fig.~\ref{fig:short1}, the significantly longer spin period 
and stronger magnetic field of the NS make the 
contribution of the luminosity released at the shock in front of the compact object comparable to the spin 
down luminosity in this regime (Eq.~\ref{eq:lxsd1}). The upper panels of Fig.~\ref{fig:short3} also show that in this case the 
magnetospheric radius and the accretion radius are relatively close one to the other. As a consequence, minor increases in the 
wind density or drops in the wind velocity cause a switch from the super-Keplerian magnetic inhibition regime to the supersonic 
propeller regime and the subsonic propeller regime, where $R_{\rm M}\ll R_{\rm a}$. A transition to the direct accretion regime 
is also observed corresponding to the largest increases in the wind density (the range spanned by the wind velocity is much more 
reduced compared to that of the density in the clumpy wind model considered here). These transitions occur, for example, around 
$t=6, 13, 15, 22, 26, 28, 35~h$. Compared to the previous two cases, it is clear that a larger magnetic field and a longer spin 
period make the more extreme accretion regimes achievable for longer amount of times. The average X-ray luminosity of the 
system is thus dramatically decreased compared to the value expected from the simple Bondi-Hoyle accretion and we approach a 
variability behavior that is more reminiscent of that of the SFXTs, with extended low emission periods separating the much 
brighter and sporadic outbursts.

\subsection{Orbital period 25.6~days}
\label{sec:long}

In this section we present similar results as those in Sect.~\ref{sec:short}, but in the case of a larger orbital 
separation between the NS and the supergiant companion. In Fig.~\ref{fig:long1}, \ref{fig:long2}, and 
\ref{fig:long3}, we show the results of the previous calculations reported in Fig.~\ref{fig:short1},\ref{fig:short2}, 
\ref{fig:short3} extended to the longer orbital period case. As it can be seen comparing each couple of figures, 
the qualitative discussion presented for the different magnetic field and spin period cases of Sec.~\ref{sec:short} 
still holds and the system in the different configurations experience similar accretion regimes. 

The main difference compared to the shorter orbital period case is that at larger distances from the 
supergiant companion, the variations in the density and velocity of the clumpy wind are much shallower. 
This leads to a substantially less pronounced variability, with X-ray outbursts becoming less and less frequent 
and low level activity intervals more and more prolonged in time. It is particularly interesting to note that 
the X-ray variability observed in the case with the higher magnetic field and longer spin period ($\mu_{33}=0.1$ and 
$P_{\rm s3}=1$, see Fig.~\ref{fig:long3}) is still the one that more closely resemble the behavior typical of SFXTs, 
with fewer and fewer prominent outbursts arising during periods 
of longer and longer low level X-ray emission.

\section{Discussion and conclusions}
\label{sec:conclusions}

In this paper we presented a first attempt to take into account the effects of the magnetic and centrifugal gating 
mechanisms during the accretion of a highly structured stellar wind onto a NS hosted in a SGXB. We showed that 
the magnetic field properties and the spin period of the compact object dramatically impact the X-ray luminosity 
released by the system. 

The non-stationary wind model we adopted in this work is well known to provide reasonably good predictions for the 
X-ray observed from massive stars. Although the density and velocity variations derived from this model might need 
to be revised in the future when 2D/3D models will be available 
(see Sect.~\ref{sec:intro}), the coupling of these variations with all the gating 
accretion regimes proposed originally by B08, allowed us to investigate how the transitions between these regimes 
can be triggered in the clumpy environment of a wind-fed SGXB. We have also been able to study which X-ray luminosity and variability 
pattern can be expected as a function of the NS spin period, magnetic field, and orbital period. 
Based on all the cases analyzed in Sect.~\ref{sec:results}, we first highlighted the fact that the introduction of the gating mechanisms 
led to a substantial reduction of the average system X-ray emission compared to that computed from a simplified Bondi-Hoyle accretion 
scenario, even in those cases in which modest magnetic field values ($\sim10^{12}$~G) and short spin periods ($\sim$10~s) are assumed 
for the compact object. This suggests that calculations of the X-ray luminosity released by NS wind-fed binaries in which the 
effect of the gating mechanisms is not taken into account could be significantly overestimated. 

We also found that the combination of a longer spin period and a higher 
magnetic field might more easily lead to an SFXT-like behavior, i.e. with sporadic bright outbursts 
emerging during periods of a much fainter X-ray emission (as previously discussed by B08). It is also worth noticing that 
the reduction of the mass accretion rate onto the compact object strongly increases with the intensity of the magnetic field, thus 
strengthening the idea that highly magnetized NSs can be hosted in SFXTs, as  
the inhibition of accretion has been convincingly identified as a key ingredient to explain the sub-luminosity of all sources in this 
class compared to classical SGXBs \citep{lutovinov13,bozzo15}. Note also that the similarity 
observed in the transitions between different accretion regimes for systems with smaller or larger orbital periods in Sect.~\ref{sec:short} and 
\ref{sec:long} is in agreement with the observational evidence that SFXTs with orbital separations spanning from a few to several days display 
remarkably similar behaviors in the X-ray domain \citep{romano14,sidoli16}.  

In general, it is difficult to reproduce the behavior of classical SGXBs within the assumption of a NS accreting from the extremely clumpy 
wind in Sect.~\ref{sec:clumpywind}. The large density and velocity variations produced by the adopted clumpy wind model lead in all cases 
to a more pronounced variability than that observed from these systems. However, it was highlighted in Sect.~\ref{sec:results} how 
placing the NS at a larger distance from the supergiant companion helps in improving the similarity 
between simulations and the observed behaviors of both classical SGXBs and SFXTs, as in these cases the stellar wind is characterized by smoother 
density and velocity contrasts. It is thus likely that, with future  
multi-dimensional stellar wind models featuring less extreme clump properties, 
the gating accretion scenario can be used to explain the difference between classical SGXBs and SFXTs by fine-tuning only 
the NS parameters.  

It has to be remarked here that, beside the limitations of the 1D clumpy wind model, the present simulations are also affected by 
a number of additional simplifications. In particular, the NS is placed in all cases at a certain distance from the supergiant 
and neither its gravitational field nor the X-rays released as a consequence of the accretion are producing any feedback 
onto the stellar wind. We know from hydrodynamics simulations of wind-fed systems that the gravitational field of the compact 
object can largely distort the accretion flow and produce significant density and velocity variations close to the compact object.  
The latter can thus enhance the X-ray variability already produced by the presence of the clumps and the 
effect of the gating mechanisms \citep{blondin91,blondin94,ManousakisWalter2014}. The photoionization of the stellar wind 
by the X-rays emitted from the accreting NS can also produce significant variations in the physical properties of the wind, 
including its composition and velocity. The most recent investigations in these respects have been presented by 
\citet{krtika12} and \citet{krtika15}. These authors have shown that the supergiant wind velocity can significantly drop 
close to the compact object or also be completely halted by the X-ray irradiation 
if the photoionization inhibits the radiative acceleration down to the surface of the supergiant star. As the accretion radius 
and the magnetospheric radius are highly sensible to variations in the wind velocity (see Eq.~\ref{eq:ra}, \ref{eq:rm1}, 
\ref{eq:rm2}, and \ref{eq:rm3}), we expect that the switches between different 
accretion regime can also be affected by the photoionization. However, calculating the reciprocal feedback between 
photoionization effects and gating accretion mechanisms in the case of a strongly in-homogeneous stellar wind  
would require the development of an extremely challenging full 3D magneto-hydrodynamical treatment of the problem. This has not been 
possible so far. As the X-ray irradiation of the stellar wind depends on the orbital separation of the system, it is clear 
that our simplified treatment of the problem would get less and less accurate for shorter orbital period systems or for binaries with 
large eccentricities. For this reason we did not consider the case of orbital periods shorter than $\sim$9~days. 
We also note that short orbital period systems ($\lesssim$3-5~days) at X-ray luminosities of  
$\lesssim$10$^{36-37}$~erg~s$^{-1}$ could also develop temporary accretion disks \citep[see, e.g.,][and refeences therein]{ducci10}, 
which effect onto the accretion process cannot be taken into account in the present version of our computations. 
We plan to include all these complications in future improved versions of the simulations presented here.

\appendix

\section{The Bondi-Hoyle accretion}
 \label{sec:appendix1}

In the Bondi-Hoyle accretion scenario adopted previously 
by \citet{oskinova2012}, the authors followed the simplified treatment 
proposed by \citet{do1973} in which a NS is traveling with a relative speed 
$v_{\rm rel}$ through a gas with density $\rho$. The mass
accretion rate is given in this case by
\begin{equation}
\dot{M}_{\rm acc}=\pi \zeta R_{\rm a}^2 v_{\rm rel} \rho_{\rm w} 
\label{eq:acrrate}
\end{equation}
where the quantities $R_{\rm a}$, $v_{\rm rel}$, and $\rho_{\rm w}$ are the 
same as those introduced in Sect.~\ref{sec:gating}. The 
factor $\zeta\sim1$ is included to take into account numerical corrections due to 
the radiation pressure and the finite cooling time of the gas. 
Combining Eq.~\ref{eq:acrrate} and \ref{eq:ra}, it can be obtained  
\begin{equation}
\dot{M}_{\rm acc}=4\pi \zeta \frac{(G M_{\rm NS})^2}{v^3_{\rm rel}}\rho_{\rm w}.  
\label{eq:sac}  
\end{equation}
In these equations, $\rho_{\rm w}$ is assumed to be the time dependent density of the wind 
provided by the clumpy wind code described in Sect.~\ref{sec:clumpywind}. 
\citet{oskinova2012} did not neglect the NS orbital velocity compared to the wind velocity, and thus 
\begin{equation}
v^2_{\rm rel}=v^2_{\rm X} + v^2_{\rm w}, 
\label{eq:vrel}
\end{equation}
where the orbital velocity of the NS, $v_{\rm X}$, is given by 
\begin{equation}
v^2_{\rm X} \approx \frac{GM_\ast}{a}.   
\label{eq:vx}
\end{equation} 
The X-ray luminosity of an accreting neutron star in this simplified scenario can thus be expressed as 
\begin{equation}
L^{\rm acc} =\eta \dot{M}_{\rm accr} c^2,
\label{eq:lx12}
\end{equation}
where $c$ is the speed of light and $\eta\sim 0.1$ is a numerical 
constant introduced to take into account unknown geometrical 
aspects of the accretion process. The value computed through Eq.~\ref{eq:lx12} 
is reported in all bottom panels of Fig.~\ref{fig:short1}, \ref{fig:short2}, \ref{fig:short3}, 
\ref{fig:long1}, \ref{fig:long2}, and \ref{fig:long3} with a solid magenta line.

\section*{Acknowledgments}
 
This publication was motivated by a team meeting sponsored by the International Space Science 
Institute at Bern, Switzerland. EB and LO thank ISSI for the financial support during their 
staying in Bern. We thank an anonymous referee for the useful comments. 

\bibliographystyle{aa}
\bibliography{sfxtclumps}

\end{document}